\documentclass[12pt,nofootinbib,floatfix,onecolumn,preprintnumbers]{revtex4}

\pdfoutput=1
\usepackage{amsmath,amssymb}
\usepackage[utf8]{inputenc}
\usepackage{graphicx}
\usepackage{cancel,float}
\usepackage{soul}
\usepackage[usenames,dvipsnames]{color}
\usepackage[colorlinks=true]{hyperref}
\usepackage{subfig}
\begin{document}
\title{Charged lepton flavor violation and electric dipole moments in the inert Zee model}
\author{Alexandra Gaviria}%
\email{alexandra.gaviria@udea.edu.co}
\author{Robinson Longas}%
\email{robinson.longas@udea.edu.co}
\author{Óscar Zapata}%
\email{oalberto.zapata@udea.edu.co}
\affiliation{
Instituto de F\'isica, Universidad de Antioquia, Calle 70 No. 52-21, A.A. 1226, Medell\'in, Colombia}
\date{\today}

\begin{abstract}
 The inert Zee model is an extension  of the Zee model for neutrino masses  to allow for a solution to the dark matter problem that involves two vector-like fields, a doublet and a singlet of $SU(2)_L$, and two scalars, also a doublet and a singlet of $SU(2)_L$, all of them being odd under an exact $Z_2$ symmetry.
  The introduction of the $Z_2$ guarantees one-loop neutrino masses, forbids tree-level Higgs-mediated flavor changing neutral currents and ensures the stability of the dark matter candidate.
  Due to the natural breaking of lepton numbers in the inert Zee model and encouraged by the ambitious experimental program designed to look for charged lepton flavor violation signals and the electron electric dipole moment, we study the phenomenology of the processes leading to these kind of signals, and establish which are the most promising experimental perspectives on that matter.
\end{abstract}

 \maketitle
\section{Introduction}
\label{sec:introduction}
Neutrino oscillations \cite{Fukuda:1998mi,Ahmad:2002jz} provide a clear evidence for lepton flavor violation (LFV) in the neutral sector, pointing out to physics beyond the Standard Model (SM). 
However, no evidence of lepton flavor violating processes in the charged sector has been found despite the great experimental effort on searching for that violation \cite{Lindner:2016bgg,Calibbi:2017uvl}.
Indeed, the experimental searches not only have reached a great sensitivity but will also be improved in the near future by, in some cases, several orders of magnitude. For instance, the MEG collaboration 
has reported an upper limit on the decay branching ratio for the rare decay $\mu\to e\gamma$ around $6\times10^{-13}$ \cite{Adam:2013mnn}, which will be improved soon by a factor of 10 \cite{Baldini:2013ke}.
Concerning the three-body decay $\mu \rightarrow 3e$, the negative searches for rare decays in the SINDRUM experiment lead to an upper limit for the branching ratio of around $10^{-12}$ \cite{Bellgardt:1987du}, 
whereas the Mu3e experiment collaboration  expects to reach the ultimate sensitivity to test such a decay in $10^{16}$ muon decays \cite{Blondel:2013ia}.
In addition, the neutrinoless $\mu$-$e$ conversion in muonic atoms is also a promising way to search for charged LFV (CLFV) signals due to the significant increase of sensitivity (up to six orders of magnitude) expected for this class of experiments \cite{Carey:2008zz,Glenzinski:2010zz,Abrams:2012er,Aoki:2010zz,Natori:2014yba,Cui:2009zz,Kuno:2013mha,Barlow:2011zza}.
Last but not least, the future plans regarding electron electric dipole moment (eEDM) \cite{Kara:2012ay,Kawall:2011zz} are also in quest for New Physics signals since  the expected sensitivity for these facilities will improve by two orders of magnitude the current bound $|d_e|<8.7\times10^{-29}e\,\cdot\,$cm~\cite{Baron:2013eja}. 
This ambitious experimental program, in turn, calls for a deep phenomenological analysis of the CLFV and EDM signals in models featuring new charged lepton interactions such as those undertaking neutrino masses.   

On the other hand, despite the abundant and compelling evidence for the massiveness of neutrinos \cite{Patrignani:2016xqp,deSalas:2017kay}, the underlying mechanism behind it remains unknown, which is not a bizarre occurrence since the particle theory responsible for the dark matter (DM) of the Universe also resists to be experimentally elucidated.
Hence, it would desirable that both phenomena may have a common origin with a New Physics laying at the electroweak scale, as happens in the radiative neutrino mass models\footnote{See Ref. \cite{Cai:2017jrq} for a review of radiative neutrino mass models.} involving a dark matter candidate at or below the TeV scale \cite{Brdar:2013iea,Restrepo:2013aga,Chen:2014ska,Ahriche:2016ixu,Simoes:2017kqb,Yao:2017vtm,Yao:2018ekp}.
Thus, the resulting model not only would constitute a way out to two of the open questions in the SM but also would have the additional bonus that the new particles may induce potentially large rates for the CLFV and EDM processes, give rise to novel observable phemomena at the LHC, and lead to signals in direct and indirect DM experiments.

In this work we consider the inert Zee model (IZM) -a dark matter realization of the Zee model for neutrino masses~\cite{Zee:1980ai,Wolfenstein:1980sy,Petcov:1982en} where the above features are present-, with the aim of pursuing a dedicated analysis of LFV processes and EDM signals\footnote{For similar dedicated studies within the context of scotogenic models see, {\it e.g.}, Refs. \cite{Toma:2013zsa,Vicente:2014wga,Chowdhury:2015sla,Cai:2016jrl,Rocha-Moran:2016enp,Klasen:2016vgl,Abada:2018zra,Esch:2018ccs}}.
As in the Zee model, neutrino masses are generated at one loop, while DM is addressed as in the inert doublet model \cite{Barbieri:2006dq,LopezHonorez:2006gr,Honorez:2010re,LopezHonorez:2010tb,Goudelis:2013uca,Garcia-Cely:2013zga,Arhrib:2013ela} since  both models share a $Z_2$-odd scalar $SU(2)_L$ doublet.
In addition, due to the new interactions, the IZM has a richer phenomenology than the minimal scotogenic model \cite{Ma:2006km}. 
Once we will have determined the viable parameter space consistent with dark matter, neutrino oscillation observables, lepton-flavor violating processes and electroweak precision tests, we will establish the most relevant experimental perspectives regarding LFV searches.
Furthermore, since the Yukawa couplings that reproduce the neutrino oscillation data are complex, which in turn constitute new sources of CP violation, we will look into the regions in the parameter space where the prospects for the eEDM are within the future experimental sensitivity \cite{Kara:2012ay,Kawall:2011zz,Baron:2013eja}.
Lastly, we will also study the possible connection, via lepton Yukawa interactions, that may exist between DM and the anomalous magnetic dipole moment of the muon \cite{Bai:2014osa,Kile:2014jea,Chen:2015jkt,Kowalska:2017iqv,Calibbi:2018rzv}.  

The paper is organized as follows: in Sec.~\ref{sec:model} we present the  generalities of the model, including neutrino masses and dark matter. In Sec.~\ref{sec:CLp} we study the LFV processes, the EDM and the magnetic dipole moment (MDM) of charged leptons. Their phenomenology is presented in Sec.~\ref{sec:numer-results-disc}. Finally, we conclude in Sec.~\ref{sec:conclusions}. 

\section{The Model}
\label{sec:model}

The new particle content of the model \cite{Restrepo:2013aga,Longas:2015sxk} consists of two vectorlike fermions, a $SU(2)_L$-singlet $\epsilon$ and a $SU(2)_L$-doublet $\Psi=(N, E)^{\text{T}}$, and two scalar multiplets, a $SU(2)_L$-singlet $S^-$ and a $SU(2)_L$-doublet $H_2=(H_2^+, H_2^0)^{\text{T}}$.
All of them are odd under the $Z_2$ symmetry, which in turn allows us to avoid Higgs-mediated flavor changing neutral currents at tree-level, forbid tree-level contributions to the neutrino masses  and render the lightest $Z_2$-odd particle stable \cite{Longas:2015sxk}. 
It follows that the most general $Z_2$-invariant Lagrangian of the model can be written as
\begin{align}
  \label{eq:Lmodel}
\mathcal{L}_{\text{IZM}}=\mathcal{L}_{\text{SM}}+\mathcal{L}_F+\mathcal{L}_S+\mathcal{L}_1+\mathcal{L}_2,
\end{align}
where $\mathcal{L}_{\text{SM}}$ is the SM Lagrangian which includes the Higgs potential $\mathcal{V}_{H_1}=\mu_1^2H_1^\dagger H_1+\lambda_1/2 (H_1^\dagger H_1)^2$, with $H_1=(0, H_1^0)^{\text{T}}$, $H_1^0=(h+v)/\sqrt{2}$, $h$ being the Higgs boson and $v=246$~GeV. $\mathcal{L}_F$ and $\mathcal{L}_S$ comprise, respectively, the  kinetic and mass terms for the new fermions, and the kinetic, mass and self-interacting terms of the new scalars,
\begin{align}
  \mathcal{L}_F&=\bar{\Psi}(i\cancel{D}-m_\Psi)\Psi+\bar{\epsilon}(i\cancel{D}-m_\epsilon)\epsilon,\\
  \mathcal{L}_S&=(D_\mu H_2)^\dagger(D^\mu H_2)-\mu_2^2H_2^\dagger H_2-\frac{\lambda_2}{2} (H_2^\dagger H_2)^2+(D_\mu S)^\dagger(D^\mu S)-\mu_S^2S^\dagger S-\lambda_S (S^\dagger S)^2.
\end{align}
The interaction terms between the scalars are included in $\mathcal{L}_1$,
\begin{align}
  -\mathcal{L}_1 & = \lambda_3 ( H_1^{\dagger}H_1 )( H_2^{\dagger}H_2 ) +  \lambda_4 ( H_1^{\dagger}H_2 )( H_2^{\dagger}H_1 )  +  \frac{\lambda_5}{2}\left[( H_1^{\dagger} H_2 )^2 + {\rm h.c.} \right] \nonumber\\
  & + \lambda_6 (S^\dagger S) (H_1^{\dagger}H_1) + \lambda_7 (S^\dagger S)(H_2^{\dagger}H_2) +  \mu  \epsilon_{ab}\left[H_1^a H_2^b S + {\rm h.c.} \right],
\end{align}
where $\epsilon_{ab}$ is the $SU(2)_L$ antisymmetric tensor with $\epsilon_{12}=1$, $H_2=(H_2^+, H_2^0)^{\text{T}}$ with $H_2^0=(H^0 + i A^0)/\sqrt{2}$,  and the scalar couplings $\lambda_5$ and $\mu$ are assumed real.
It is worth mentioning that $H_2^0$ does not develop a vacuum expectation value in order to ensure the conservation of the $Z_2$ symmetry. 
Note that the scalar potential is rather similar to one of the singlet-doublet scalar DM model \cite{Kadastik:2009dj,Kadastik:2009cu,Kakizaki:2016dza,Liu:2017gfg} or two Higgs doublet models plus a scalar singlet, see {\it e.g.} Refs.~\cite{Chen:2013jvg,Muhlleitner:2016mzt}, with the main difference that our $SU(2)_L$-singlet  scalar is electrically charged (which in turn implies that $S$ has also a null vacuum expectation value and that the charged scalars get mixed instead of neutral ones).  
Finally, $\mathcal{L}_2$ includes the new Yukawa interaction terms\footnote{Here we assume parity conservation for the new sector, and thus we  neglect the $\bar{\Psi}\gamma_5H_1\epsilon$ term.}:
\begin{align}
  \label{eq:YukawaLagrangian}
  -\mathcal{L}_2&=  \eta_i\bar{L}_{i}H_2\epsilon + \rho_i \bar{\Psi}H_2 e_{Ri} + y \bar{\Psi} H_1 \epsilon 
                  + f^*_i \overline{L^c_{i}} \Psi S^+  + {\rm h.c},
\end{align}
where  $L_i$ and $e_{Ri}$ ($i=1,2,3)$ are the SM leptons, doublets and singlets of $SU(2)_L$, respectively.  $\eta_i$, $\rho_i$ and $f_i$ are Yukawa couplings controlling the new lepton interactions, while $y$ leads to the mixing among the $Z_2$-odd charged fermions.
Note that only $\eta_i$ and $f_i$ involve neutrino interactions, so they are the ones participating in the neutrino mass generation.

After electroweak symmetry breaking, the $Z_2$-odd scalar spectrum consists of a CP-even state $H^0$ and a CP-odd state $A^0$, and two charged states $\kappa_{1,2}$. Their masses are given by 
\begin{align}
  m^2_{H^0,A^0}&=\mu_{2}^{2}+\frac{1}{2}\left(\lambda_{3}+\lambda_{4}\pm\lambda_{5}\right)v^2,\\
  m_{\kappa_1,\kappa_2}^2&=\frac{1}{2}\left\{m_{H^{\pm}}^2 + m_{S^\pm}^2 \mp [( m_{H^{\pm}}^2 - m_{S^\pm}^2 )^2 + 2\mu^2v^2]^{1/2}\right\},
\end{align}
where $m_{H^{\pm}}^2 = \mu_2^2 + \frac{1}{2}\lambda_3 v^2$ and $m_{S^{\pm}}^2 = \mu_S^2 + \frac{1}{2}\lambda_6 v^2$.
The scalar mixing angle $\delta$ is defined through $\sin{2\delta}=(\sqrt{2} \mu v)/(m_{\kappa_2}^2-m_{\kappa_1}^2)$. 
On the other hand, the $Z_2$-odd fermion spectrum involves two  charged fermions $\chi_{1,2}$ with
\begin{align}
  m_{\chi_{1,2}} &=   \frac{1}{2}\left\{ m_{\Psi} + m_{\epsilon} \mp [( m_{\Psi} - m_{\epsilon} )^2 + 2y^2v^2]^{1/2}\right\},
  \end{align}
and a mixing angle given by  $\sin{2\alpha} = (\sqrt{2} y v)/(m_{\chi_2} - m_{\chi_1})$, along with the neutral Dirac fermion $N$, with a mass $m_N = m_{\Psi}$ fulfilling $m_{\chi_1}\leq m_N\leq m_{\chi_2}$.

With respect to DM in the IZM, $H^0$ is the DM candidate\footnote{Without loss of generality we assume $H^0$ to be the DM candidate. Note also that the neutral fermion $N$ can not play the role of the DM candidate since $m_{\chi_1}\leq m_N$.} as long as it remains as the lightest $Z_2$-odd  particle in the spectrum.
Hence, we expect the DM phenomenology to be similar to the one in the inert doublet model (IDM) in scenarios where the particles not belonging to the IDM ($\kappa_{1,2}$, $\chi_{1,2}$ and $N$) do not participate in the DM annihilation processes \cite{Longas:2015sxk}.  
Accordingly, the viable DM mass range for this scenario (the same of the one in the IDM) is divided into two regimes \cite{Barbieri:2006dq,LopezHonorez:2006gr,Honorez:2010re,LopezHonorez:2010tb,Goudelis:2013uca,Garcia-Cely:2013zga,Arhrib:2013ela}: the low mass regime, $m_{H^0} \simeq m_{h}/2$, and the high mass regime, $m_{H^0}\gtrsim 500$ GeV. 
Since in the latter regime the CLFV processes are quite suppressed (the corresponding rates scale as $m_{\chi_\omega}^{-4}$ or $m_{\kappa_\beta}^{-4}$), in our numerical analysis  we will only consider the low mass regime. 

\begin{figure}[t!]
\includegraphics[scale=0.7]{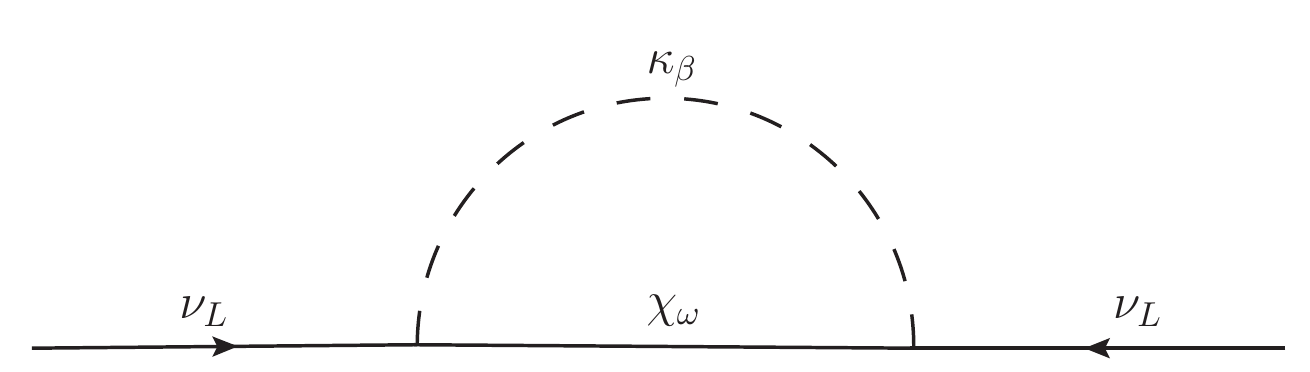}
\caption{One-loop diagram leading to  neutrino Majorana masses. }
\label{fig:neutrinomass}
\end{figure}

In this model the neutrino masses are generated at one-loop thanks to the scalar and fermion mixings and to the Yukawa interactions mediated by $\eta_i$ and $f_i$. 
From Fig.~\ref{fig:neutrinomass}, the  neutrino mass matrix in the mass eigenstates is given by 
\begin{align}\label{eq:neutrinomass}
  [M_{\nu}]_{ij} &= \zeta [\eta_i f_j + \eta_j f_i ],
\end{align}
where $\zeta =(\sin2\alpha \sin2\delta)/(64 \pi^2)\sum_nc_nm_{\chi_n}I(m_{\kappa_1}^2,m_{\kappa_2}^2,m_{\chi_n}^2)$,  $c_1=-1$, $c_2=+1$ and $I(a,b,c)= b\ln(b/c)/(b-c) -  a\ln(a/c)/(a-c)$.
Note that for a vanishing scalar or fermion mixing the neutrino masses are zero and that, due to the flavor structure of $M_{\nu}$, the lightest neutrino is massless. 
This, the masslessness of the lightest neutrino, entails several phenomenological consequences: $i)$ there is only single Majorana CP phase since the second phase can be absorbed by a redefinition of the massless neutrino field; $ii)$ the two remaining neutrino masses are entirely set by the solar and atmospheric mass scales: for normal hierarchy (NH) $m_1=0$, $m_2=\sqrt{\Delta m_{\text{sol}}^2}$ and $m_3=\sqrt{\Delta m_{\text{atm}}^2}$, while for inverted hierarchy (IH) $m_1=\sqrt{\Delta m_{\text{atm}}^2}$, $m_2=\sqrt{\Delta m_{\text{sol}}^2+m_1^2}\approx\sqrt{\Delta m_{\text{atm}}^2}$ and $m_3=0$; 
$iii)$ the amplitude for neutrinoless double beta decay \cite{Rodejohann:2011vc} presents a lower bound, which for the case of IH lies within the sensitivity of future facilities dedicated for that goal \cite{Reig:2018ztc}. 

In Ref.~\cite{Longas:2015sxk} it was shown that, using Eq.~(\ref{eq:neutrinomass}) and the diagonalization condition\footnote{We work in the basis where the charged-lepton Yukawa matrix is diagonal.} $U^{\text{T}}M_{\nu}U ={\rm diag}(m_1,m_2,m_3)$ with  $U=VP$ and $P=\mbox{diag}(1,e^{i\phi/2},1)$~\cite{Agashe:2014kda}, it is possible to express five of the six Yukawa couplings $\eta_i$ and $f_i$ in terms of the neutrino low energy observables. 
Consequently, the most general Yukawa couplings that are compatible with the neutrino oscillation data are given by
\begin{align}\label{eq:yuks-f-eta}
  &\eta_i=|\eta_1|\frac{A_i}{\beta_{11}}, \hspace{1cm}f_i=\frac{1}{2\zeta}\frac{\beta_{ii}}{\eta_i},
\end{align}
where 
\begin{align}
  \beta_{ij} &= e^{i\phi} m_2V_{i2}^*V_{j2}^*+ m_3V_{i3}^*V_{j3}^*,\nonumber\\
  A_{j} &= \pm\sqrt{-e^{i\phi} m_2m_3(V_{12}^*V_{j3}^*-V_{13}^*V_{j2}^*)^2e^{i2\text{Arg}(\eta_1)}}+\beta_{1j}e^{i\text{Arg}(\eta_1)},\,\,\,  \mbox{for NH}, \label{eq:betaij1}\\
  \beta_{ij} &= m_1V_{i1}^*V_{j1}^*+ e^{i\phi} m_2V_{i2}^*V_{j2}^*,\nonumber\\ A_{j} &= \pm\sqrt{-e^{i\phi} m_1m_2(V_{11}^*V_{j2}^*-V_{12}^*V_{j1}^*)^2e^{i2\text{Arg}(\eta_1)}}+\beta_{1j}e^{i\text{Arg}(\eta_1)},\,\,\, \mbox{for IH}.\label{eq:betaij2} 
\end{align}
In this way, it is always possible to correctly reproduce the neutrino oscillation parameters in the present model\footnote{This result is also valid for models with a neutrino mass matrix having the same flavor structure of $M_\nu$ in Eq.~(\ref{eq:neutrinomass}).}.

\section{Charged lepton processes}\label{sec:CLp}
\begin{figure}[t]
    \includegraphics[scale=0.8]{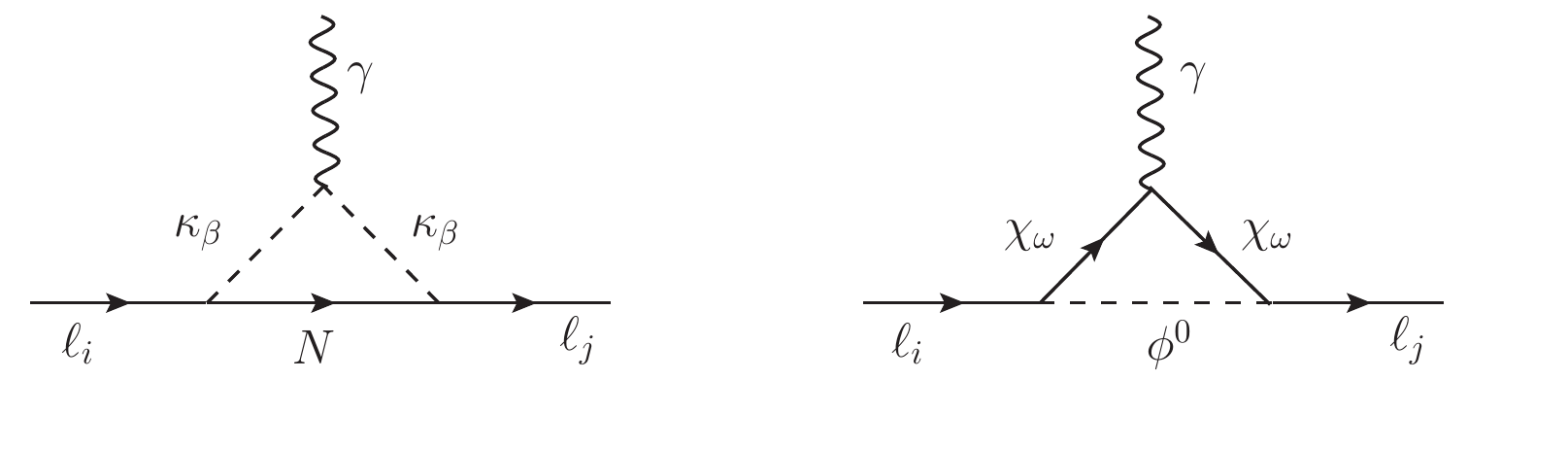}
    \caption{One-loop diagrams leading to eEDM, anomalous muon MDM, and $\ell_i \rightarrow \ell_j \gamma$ decays when $i=j=1$, $i=j=2$ and $i\neq j$, respectively. Here $\phi^0$ denotes the two $Z_2$-odd neutral scalars $A^0$ and $H^0$.}
    \label{fig:muegammadiagrams}
\end{figure}
Once lepton flavor violation is allowed via neutrino Majorana masses, LFV processes involving charged leptons such as $\ell_{i} \rightarrow \ell_{j}\gamma$, $\ell_{i} \rightarrow 3\ell_{j}$ and $\mu-e$ conversion in nuclei are unavoidable.
In the IZM model such processes are generated at one-loop level involving the $\eta_{i}$, $f_{i}$ and $\rho_{i}$ Yukawa interactions (see Eq. (\ref{eq:YukawaLagrangian})), which in turn implies that they are mediated by, both charged and neutral, $Z_2$-odd fermions and scalars. 
It is worth mentioning that $\rho_i$ may enhance the rates for the CLFV processes and EDMs since it is not subject to the neutrino oscillation constraints (only $\eta_i$ and $f_i$ enter in the neutrino mass generation).

The triangle diagrams leading to $\ell_{i} \rightarrow \ell_{j} \gamma$ processes are displayed in Fig.~\ref{fig:muegammadiagrams}. The corresponding branching ratios (neglecting lepton masses at final states) are given by
\begin{align}
\label{eq:bramuegamma}
 \mathcal{B}\left(\ell_{i} \rightarrow \ell_{j} \gamma \right)
 = \frac{3\alpha_{em}}{64\pi m_{\mu}^2G_F^2} \left( \left|\Sigma_L\right|^2 + \left|\Sigma_R\right|^2 \right) \mathcal{B}\left(\ell_{i} \rightarrow \ell_{j} \nu_i \bar{\nu_j} \right),
\end{align}
where $\alpha_{em}$ is the electromagnetic fine structure constant, $G_F$ is the Fermi constant and $\Sigma_L$, $\Sigma_R$ are given by
\begin{align}\label{eq:sigmaL}
  \Sigma_L =&  - \eta_{i}^* \rho_{j}^* s_\alpha c_\alpha\left[m_{\chi_1} \mathcal{G}_1(m^2_{\chi_1},m_{A^0}^2,m_{H^0}^2)-m_{\chi_2}\mathcal{G}_1(m^2_{\chi_2},m_{A^0}^2,m_{H^0}^2)\right] \nonumber\\
	    & - m_{\ell_i}  \rho_{i}^* \rho_{j}\left[s^2_\alpha\mathcal{F}_1(m^2_{\chi_2},m_{A^0}^2,m_{H^0}^2)+c^2_\alpha\mathcal{F}_1(m^2_{\chi_1},m_{A^0}^2,m_{H^0}^2)\right]\nonumber\\
            &+m_{\ell_i} \rho_{i}^* \rho_{j} \left[c^2_\delta\mathcal{F}_2(m^2_{\kappa_1},m_{N}^2)+s^2_\delta\mathcal{F}_2(m^2_{\kappa_2},m_{N}^2)\right] \;, \\
  \Sigma_R =&  - \rho_{i} \eta_{j} s_\alpha c_\alpha\left[m_{\chi_1} \mathcal{G}_1(m^2_{\chi_1},m_{A^0}^2,m_{H^0}^2)-m_{\chi_2}\mathcal{G}_1(m^2_{\chi_2},m_{A^0}^2,m_{H^0}^2)\right] \nonumber \\
            & -m_{\ell_i} \eta_{i}^*\eta_{j} \left[c^2_\alpha\mathcal{F}_1(m^2_{\chi_2},m_{A^0}^2,m_{H^0}^2) +s^2_\alpha\mathcal{F}_1(m^2_{\chi_1},m_{A^0}^2,m_{H^0}^2)\right]\nonumber\\
            &+m_{\ell_i}f_{i}^*f_{j} \left[s^2_\delta\mathcal{F}_2(m^2_{\kappa_1},m_{N}^2)+c^2_\delta\mathcal{F}_2(m^2_{\kappa_2},m_{N}^2)\right]\;\label{eq:sigmaR}.
\end{align}
The loop functions  are reported in the Appendix. 
Note that, in contrast to neutrino masses, these branching ratios are not double suppressed by the mixing of $Z_2$-odd particles.
This implies that, as is expected, the rare decays $\ell_{i} \rightarrow \ell_{j} \gamma$ do not depend on  whether neutrino masses are zero or not.
On the other hand, note that the IZM has an additional contribution (right diagram) to $\ell_{i} \rightarrow \ell_{j} \gamma$ with respect to the minimal scotogenic model \cite{Ma:2006km}. Indeed, in that diagram precisely enter $\rho_i$, the Yukawa couplings that are not affected by neutrino oscillation constraints. 

\begin{figure}[t]
\includegraphics[scale=0.8]{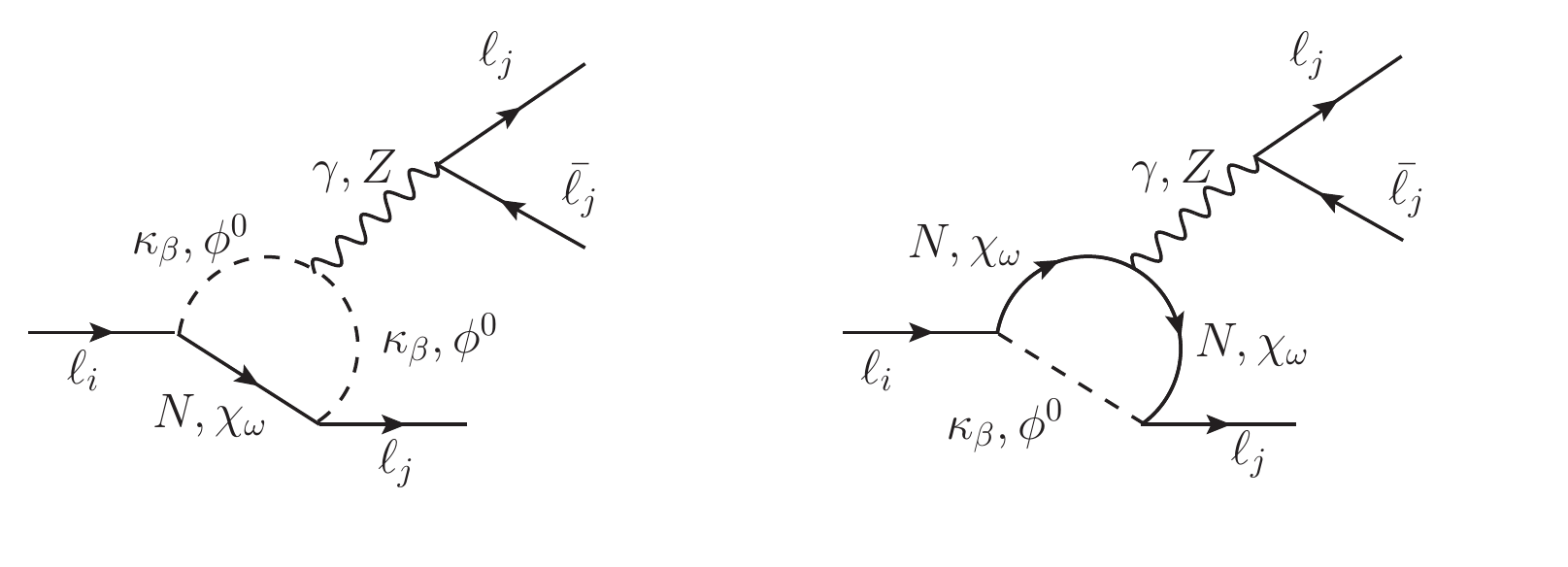}\\
\includegraphics[scale=0.8]{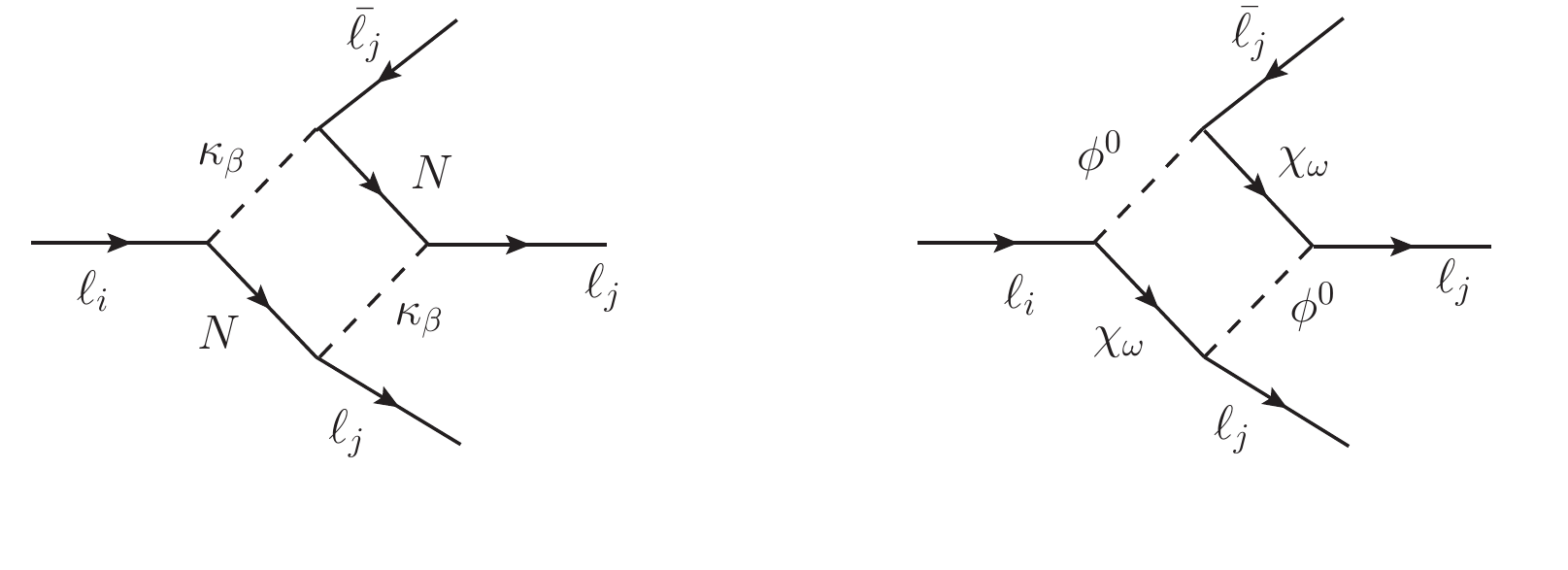}
\caption{One-loop diagrams contributing to $\ell_i \rightarrow \ell_j\bar{\ell_j}\ell_j$ decays, where $\phi^0$ denotes the two $Z_2$-odd neutral scalars $A^0$ and $H^0$. }
\label{fig:muegammadiagrams2}
\end{figure}

Concerning the $\ell_i \rightarrow \ell_j\bar{\ell_j}\ell_j$ there are two class of diagrams (see Fig.~\ref{fig:muegammadiagrams2}) leading to such processes: the $\gamma$- and $Z$- penguin diagrams (top panels) and the box diagrams (bottom panels).   
There is also a contribution from Higgs-penguin diagrams which, nevertheless, is suppressed for the first two charged leptons generations due to their small Yukawa couplings.
The contribution of those processes involving tau leptons is not negligible but the corresponding limits are less restrictive. 
Therefore, the $\ell_i \rightarrow \ell_j\bar{\ell_j}\ell_j$ processes contain four kind of contributions: the photonic monopole, photonic dipole,  $Z$-penguin and boxes. 
In contrast, the photonic dipole contribution is the only one present in  $\ell_{i} \rightarrow \ell_{j} \gamma$ processes.

Finally, the $\mu-e$ conversion diagrams are obtained when the pair of lepton lines attached to the photon and $Z$ boson in the penguin diagrams (see top panels of Fig.~\ref{fig:muegammadiagrams2}) are replaced by a pair of light quark lines\footnote{Higgs-penguin diagrams are again suppressed, in this case by the Yukawa couplings of light quarks.}. There are no box diagrams since the $Z_2$-odd particles do not couple to quarks at tree level. 
Accordingly, the photonic non-dipole and dipole terms along the $Z-$penguin are the only terms that contribute to the $\mu-e$ conversion processes. 
Since the calculations of the $\ell_i \rightarrow \ell_j\bar{\ell_j}\ell_j$ and $\mu-e$ rates are quite involved, we implement the {\tt FlavorKit} code \cite{Porod:2014xia}  to use the corresponding full expressions.

Since the parameters $\beta_{ij}$ and $A_j$ in Eqs. (\ref{eq:betaij1}) and (\ref{eq:betaij2}) are in general complex, it follows that the Yukawa couplings required for  neutrino oscillation data constitute new sources of CP violation in the lepton sector\footnote{The Yukawa couplings become real in a CP-conserving scenario with $\lambda=-1$ and $\eta_1$ real \cite{Longas:2015sxk}.}.
Thus the IZM brings with it new contributions to the EDM of charged leptons with the distinctive feature that the leading contribution arises at one-loop level (see right top diagram of Fig.~$\ref{fig:muegammadiagrams}$ with $i=j=1$), in constrast with the one obtained in the minimal scotogenic model where it arises a two loops \cite{Abada:2018zra}.
Accordingly, large values for the eEDM may be expected. 
The analytical expression reads 
\begin{equation} \label{eq:EDM}
 d_{\ell_i} = \frac{e}{2^6 \pi^2} s_{2\alpha} \operatorname{Im}(\rho_i \eta_i) \left[m_{\chi_2}\mathcal{I}_1( m^2_{\chi_2}, m_{H^0}^2, m_{A^0}^2 )-
 m_{\chi_1}\mathcal{I}_1( m^2_{\chi_1}, m_{H^0}^2, m_{A^0}^2 )\right],
\end{equation}
where the loop function is given in the Appendix.
It is clear then that $d_{\ell_i}$ becomes suppressed for small fermion mixing angles.  
On the other hand, note that the left diagram of Fig.~$\ref{fig:muegammadiagrams}$ does not contribute to the EDM because the neutral Dirac fermion, $N$, only has one single chiral coupling to either $\ell_i$ or $\ell^c_i$, {\it i.e.}, such a diagram conserves CP. 

Yukawa interactions also lead to contributions to anomalous magnetic dipole moments of the charged leptons. 
These contributions can be cast as 
\begin{equation}
\label{eq:CAMM}
  a_{\ell_i}^{\text{IZM}}= a_{\ell_i}^\mathrm{F} + a_{\ell_i}^\mathrm{S},
\end{equation}
where 
\begin{align}
  a_{\ell_i}^\mathrm{F} & = \frac{m_{\ell_i}^2}{2^5 \pi^2}
                       \Big[\left(|\rho_i|^2 c^2_\alpha +
                       |\eta_i|^2 s^2_\alpha\right)  
                       \mathcal{H}_1 (m_{\chi_1}^2, m_{H^0}^2,m_{A^0}^2 )
                       +\left(|\rho_i|^2 s^2_\alpha +  
                       |\eta_i|^2 c^2_\alpha\right) \mathcal{H}_1
                       (m_{\chi_2}^2, m_{H^0}^2,m_{A^0}^2 ) \Big. \nonumber
  \\ &  \hspace{0.3 cm}+ \frac{1}{m_{\ell_i}}\Big. s_{2\alpha} \operatorname{Re}(\rho_i\eta_i)       [ m_{\chi_1} \mathcal{I}_1 (m_{\chi_1}^2, m_{H^0}^2,m_{A^0}^2 ) -
       m_{\chi_2} \mathcal{I}_1 (m_{\chi_2}^2, m_{H^0}^2,m_{A^0}^2 )] \Big], \label{g-2-1}\\ 
  a_{\ell_i}^\mathrm{S}& = - \frac{m_{\ell_i}^2}{2^4 \pi^2} 
                      \Big[[|\rho_i|^2 c^2_\delta + |f_i|^2 s^2_\delta] \mathcal{H}_2 (m_{\kappa_1}^2, m_{N}^2 )  +
                      [|\rho_i|^2 s^2_\delta + |f_i|^2 c^2_\delta] \mathcal{H}_2 (m_{\kappa_2}^2, m_{N}^2 ) \Big], \label{g-2-2}
\end{align}
with $a_{\ell_i}^\mathrm{F}$  being the contribution involving  charged fermions and  neutral scalars,  and $a_{\ell_i}^\mathrm{S}$  the  contribution involving charged scalars and neutral fermions. 

\section{Results and discussion}
\label{sec:numer-results-disc}
\begin{table}[t!]
\begin{center}
\begin{tabular}{|cc|cc|c|}
\hline 
Observable & & Present limit & & Future sensitivity \\
\hline\hline 
$\mathcal{B}(\mu \to e\gamma)$ & & $5.3 \times 10^{-13}$ \cite{Adam:2013mnn} & & $ 6.3 \times 10^{-14}$ \cite{Baldini:2013ke} \\
$\mathcal{B}(\tau \to e\gamma)$&&$3.3 \times 10^{-8}$ \cite{Aubert:2009ag,Bona:2007qt,Miyazaki:2012mx}&& $3\times 10^{-9}$ \cite{Aushev:2010bq}\\
$\mathcal{B}(\tau\to\mu\gamma)$&& $4.4 \times 10^{-8}$ \cite{Aubert:2009ag,Bona:2007qt,Miyazaki:2012mx}&& $3\times 10^{-9}$ \cite{Aushev:2010bq}\\
$\mathcal{B}(\mu \to eee)$ & & $ 1.0\times 10^{-12}$ \cite{Bellgardt:1987du} & &$   10^{-16}$  \cite{Blondel:2013ia}   \\  
$\mathcal{B}(\tau \to eee)$ & & $ 4.4\times 10^{-8}$ \cite{Hayasaka:2010np} & &$  3\times 10^{-9}$  \cite{Aushev:2010bq}  \\  
$\mathcal{B}(\tau \to \mu\mu\mu)$ & & $ 2.1\times 10^{-8}$ \cite{Hayasaka:2010np} & &$   10^{-9}$  \cite{Aushev:2010bq}  \\  
${\rm R_{\mu e}({\rm Ti})}$ & & $4.3 \times 10^{-12}$ \cite{Dohmen:1993mp} & & $  10^{-18}$ \cite{Abrams:2012er} \\
${\rm R_{\mu e}({\rm Au})}$ & & $7.3 \times 10^{-13}$ \cite{Bertl:2006up} & &  $-$ \\
\hline
\end{tabular}
\end{center}
\caption{ Current bounds and projected sensitivities for CLFV observables.}
\label{tab:LFVprocessesexperimetal}
\end{table}

In order to obtain the particle spectrum and low energy observables we have used {\tt SPheno} \cite{Porod:2003um,Porod:2011nf} and the {\tt FlavorKit} \cite{Porod:2014xia} of {\tt SARAH} \cite{Staub:2013tta,Staub:2015kfa}, and {\tt MicrOMEGAS} \cite{Belanger:2013oya} to calculate the DM relic abundance.
The set of free parameters of the model relevant for our analysis  has been varied as\footnote{The expressions for the relations between some of the scalar potential parameters and the scalar masses are given in Ref. \cite{Longas:2015sxk}.} 
\begin{align}
  \label{eq:scanLFVlow2}
  & 10^{-5} \leq | \eta_1 |,|\rho_1|,|\rho_2|,|\rho_3| \leq 3\;;\;\nonumber\\
  &0 \leq \text{Arg}(\eta_1), \text{Arg}(\rho_1), \text{Arg}(\rho_2), \text{Arg}(\rho_3)\leq2\pi\;;\nonumber\\
  &-\pi/2 \leq \alpha, \delta \leq \pi/2\;;\;    \nonumber\\
  & 100 \, {\rm GeV} \leq m_{A^0},\, m_{\kappa_1},m_{\chi_1}  \leq 500\, {\rm GeV}\;;\;\nonumber\\
  &m_{\kappa_2} =  [m_{\kappa_1},600\,{\rm GeV}]\, 
    \;;\; m_{\chi_2} =  [m_{\chi_1},600\, {\rm GeV}].
\end{align}
Since we perform the numerical analysis only for the low DM regime, the mass of the DM candidate has been fixed at  $m_{H^0}=60$ GeV and the scalar coupling $\lambda_L\sim3\times10^{-4}$, in order to reproduce the DM relic density measurement reported by the Planck collaboration~\cite{Ade:2015xua}\footnote{We have checked that the variation of $m_{H^0}$ within the allowed mass range does not significantly affect the charged lepton observables.}. 
We also ensure that the extra coannihilation processes do not modify the expected DM relic density and that the $S$, $T$ and $U$ oblique parameters remain within the $3\sigma$ level \cite{Baak:2014ora}. The LEP II constraints on the masses of the charged $Z_2$-odd fermions and scalars \cite{Achard:2001qw,Lundstrom:2008ai} are also automatically taken into account in the scan definition.
The limits from LHC Run I dilepton searches do not further constrain the parameter space under study since those apply for low scalar masses \cite{Belanger:2015kga,Ilnicka:2015jba}.
However, since the contribution of the $Z_2$-odd charged scalars and fermions to the Higgs diphoton decay may induce large deviations from the LHC Run~2 measurement \cite{Aaboud:2018xdt,Sirunyan:2018ouh}, we discard those benchmark points that deviate beyond $2\sigma$ from the central value reported by the CMS and ATLAS collaborations.
%
Regarding the low energy neutrino parameters we consider both normal and inverted hierarchies for the neutrino mass spectrum and take the current best fit values reported in Ref.~\cite{deSalas:2017kay}. This, along with the scan values, allows us to calculate the set of remaining Yukawa couplings through Eq.~(\ref{eq:yuks-f-eta}). For this set of couplings and for the fermion mixing parameter we assume the following constraints $10^{-5}\leq|\eta_2|,|\eta_3|,|f_1|,|f_2|,|f_3|\leq3$ and $|y|<3$.

\begin{figure}[t!]
  \includegraphics[scale=0.5]{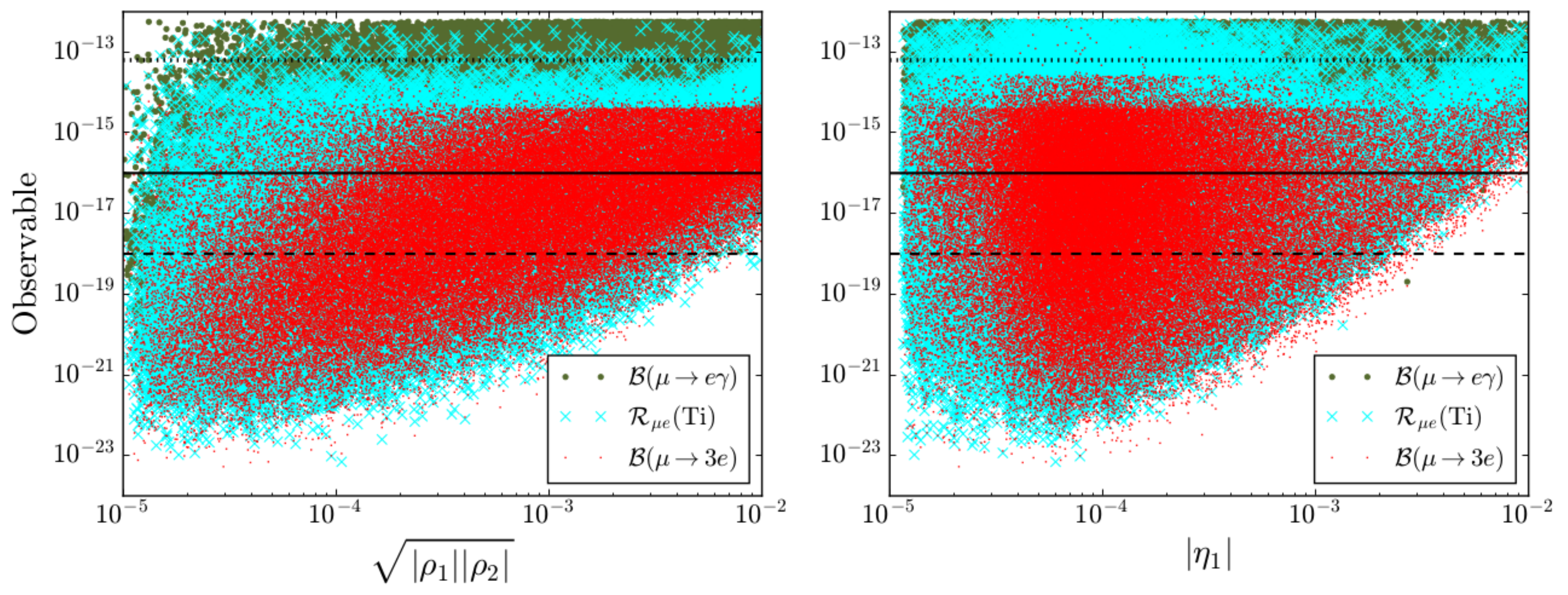}
  \includegraphics[scale=0.5]{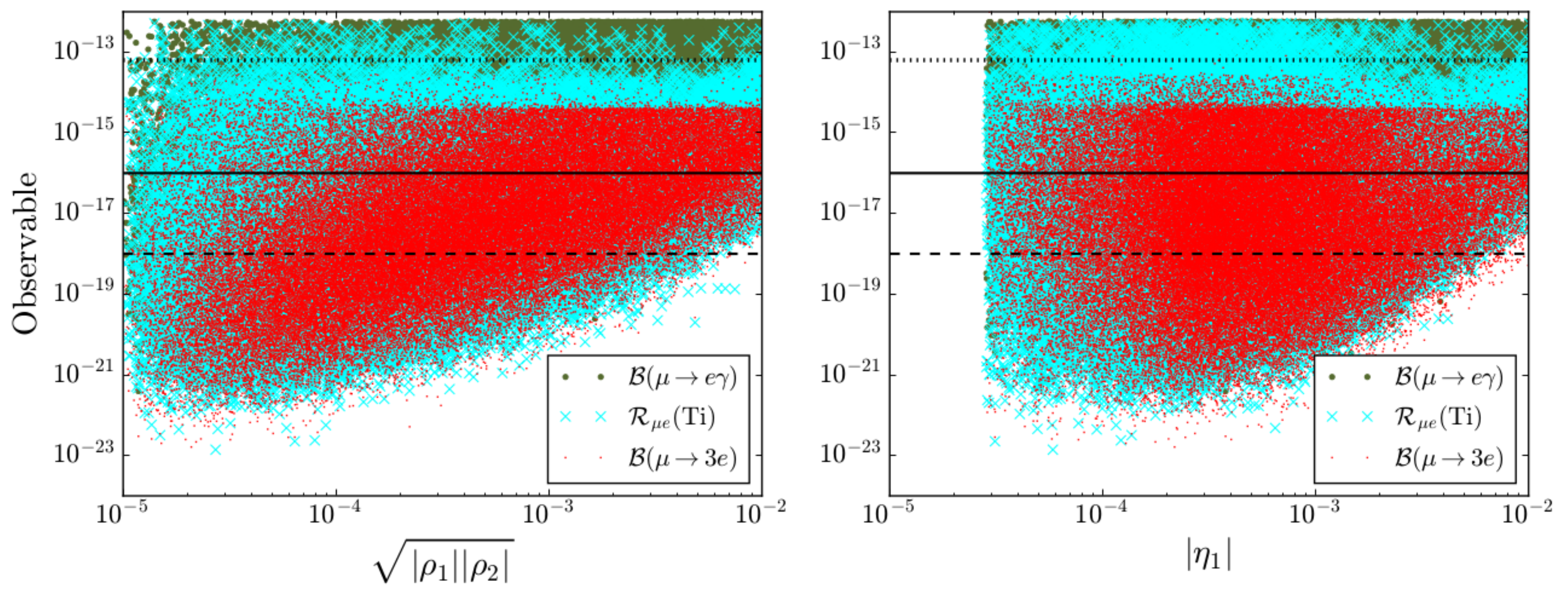}
  \caption{Rates for CLFV processes involving muons as a function of $|\eta_1|$ (right) and $\sqrt{|\rho_1||\rho_2|}$ (left panel). The upper (lower) panels are for a normal (inverted) hierarchy in the neutrino spectrum.  The dotted, solid and dashed horizontal  lines represent the sensitivity limit expected for the future searches for $\mathcal{B}(\mu\rightarrow e \gamma)$, $\mathcal{B}(\mu\rightarrow 3e)$ and $\mathcal{R}_{\mu e}$, respectively.}
  \label{fig:LFVNHIH}
\end{figure}

With respect to the CLFV observables we consider the current experimental bounds and their future expectations shown in Table \ref{tab:LFVprocessesexperimetal}, while for EDMs we have taken into account the current experimental limits $|d_e|\leqslant 8.7\times 10^{-29}{\rm \,e\,cm}$ \cite{Baron:2013eja}, $|d_{\mu}|\leqslant 1.9\times 10^{-19}{\rm \,e\,cm}$ \cite{Bennett:2008dy} and $|d_{\tau}|\leqslant 4.5\times 10^{-19}{\rm \,e\,cm}$ \cite{Inami:2002ah}.
Note that the strongest limit is set for the eEDM, and for that reason we do not display the muon and tau EDM results\footnote{We have verified that these contributions are below the current limits.}.
In regard to future eEDM searches, it is worthwhile mentioning that the ACME collaboration will increase by two orders of magnitude the current bound \cite{Baron:2013eja,futureACME}.
  
Lastly, the parameters entering in the scalar potential are subject to the following perturbativity and vacuum stability constraints\footnote{The scalar mixing parameter, $\mu$, has been chosen to lie around or below the electroweak scale in order to avoid fine-tuned cancellations in the scalar masses.}:
\begin{align}
  \label{eq:contraints_scalarparameters}
  &\mu_1^2<0,\;\; \mu_2^2,\,\mu_S^2>0,\;\;|\mu|<500\,\text{GeV},\;\; 
  \lambda_1\mu_2^2>\left(\lambda_3+\lambda_4\pm|\lambda_5|\right)\mu_1^2,\;\;|\lambda_S|,\,|\lambda_i| < 4\pi \;, \nonumber \\
  & \lambda_1,\,\lambda_2,\,\lambda_S > 0\;,\;\;   \lambda_6 > - \sqrt{\frac{\lambda_1 \lambda_S}{2}}\;,\;\; 
   \lambda_7 > - \sqrt{\frac{\lambda_2 \lambda_S}{2}} \;,\;\;
  \lambda_3 + \lambda_4 - | \lambda_5 | + \sqrt{\lambda_1 \lambda_2}> 0\;.
\end{align}
Furthermore, we also impose the upper limits set by the perturbative unitarity of the $S$-matrix \cite{Ginzburg:2005dt,Muhlleitner:2016mzt,Liu:2017gfg},
\begin{align}\label{eq:unitarity}
&|\lambda_3\pm\lambda_4|\leq8\pi, \,\,\, |\lambda_3\pm\lambda_5|\leq8\pi, \,\,\,| \lambda_3+2\lambda_4\pm3\lambda_5|\leq8\pi, \,\,\, |\Lambda_1|,|\Lambda_2|,|\Lambda_3|\leq8\pi,\nonumber\\
  &\frac{1}{2}\left|\lambda_1+\lambda_2\pm\sqrt{(\lambda_1-\lambda_2)^2+4\lambda_4^2}\right|\leq8\pi, \,\,\, \frac{1}{2}\left|\lambda_1+\lambda_2\pm\sqrt{(\lambda_1-\lambda_2)^2+4\lambda_5^2}\right|\leq8\pi.
\end{align}
Here $\Lambda_i$ correspond to the three real eigenvalues of the matrix
\begin{align}
\begin{pmatrix}
 3\lambda_1 & 2\lambda_3+\lambda_4& \sqrt{2}\lambda_6\\
 2\lambda_3+\lambda_4 &  3\lambda_2 & \sqrt{2}\lambda_7\\
 \sqrt{2}\lambda_6 & \sqrt{2}\lambda_7 & \lambda_S 
\end{pmatrix}.
\end{align}
All these theoretical conditions constrain the mass splittings among the $Z_2$-odd scalar particles. 

\begin{figure}[t!]
  \includegraphics[scale=0.45]{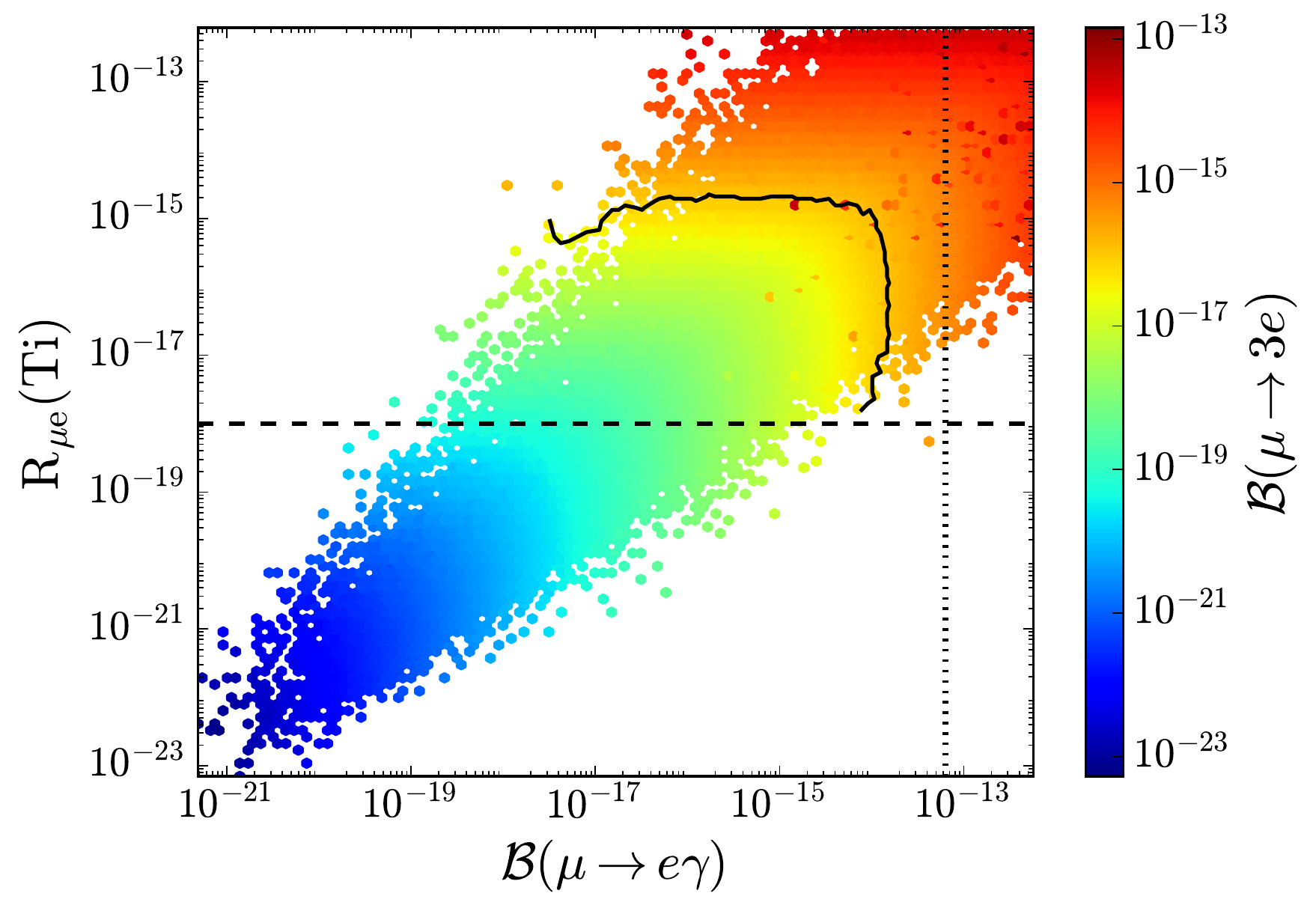}
  \includegraphics[scale=0.45]{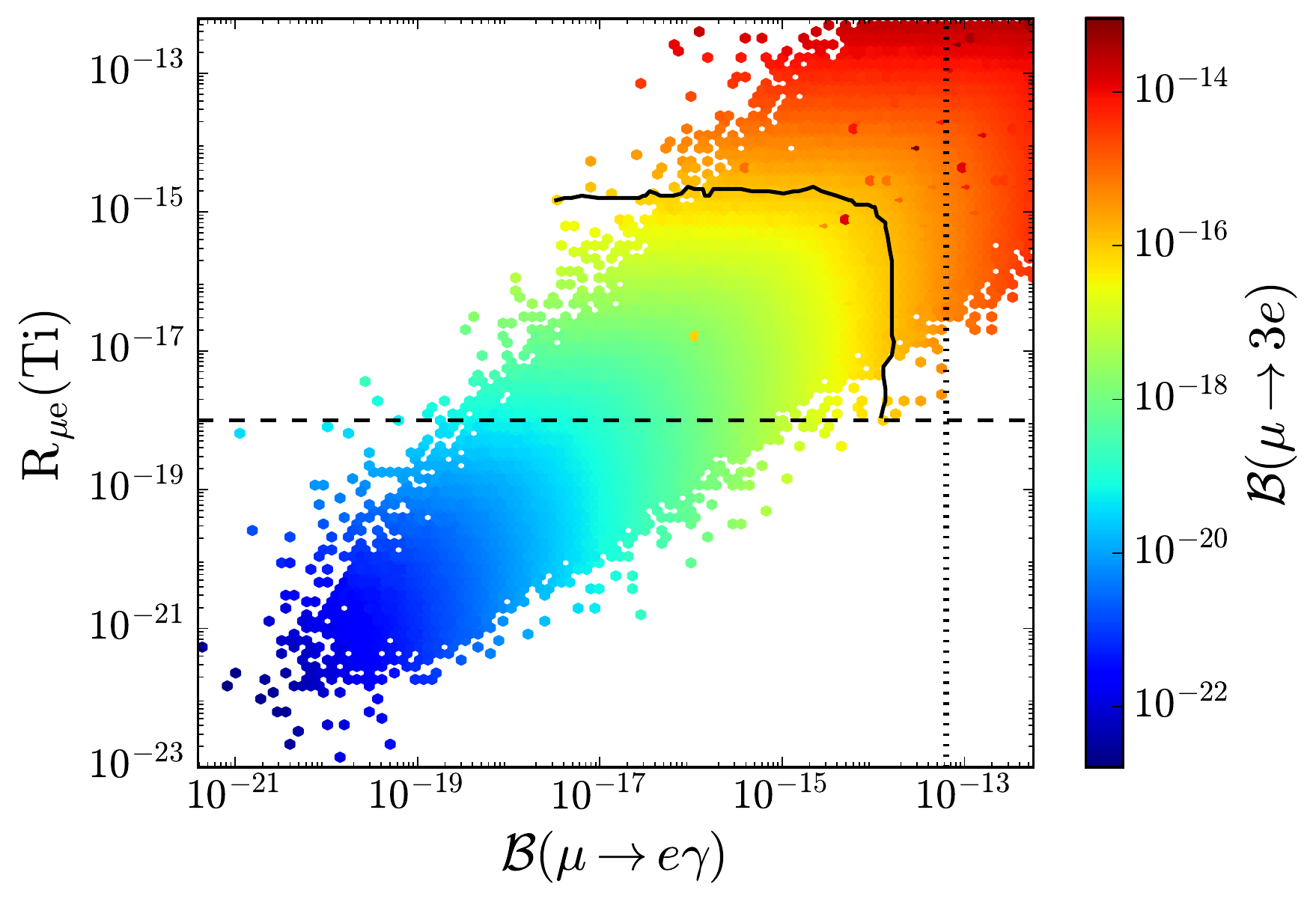}
  \caption{The available parameter space of the IZM and the future exclusion zones coming from the most constraining CLFV searches. Left (right) panel is for NH (IH). The dotted, solid and dashed lines represent the expected sensitivity for the future searches regarding $\mathcal{B}(\mu\rightarrow e \gamma)$, $\mathcal{B}(\mu\rightarrow 3e)$ and $\mathcal{R}_{\mu e}$, respectively.}
  \label{fig:LFV1}
\end{figure}

In the following we present the results for the viable benchmark points obtained from the scan.
Only the results for CLFV processes involving muon leptons in the inital state are displayed since the ones with tau leptons are out the reach of the forthcoming facilities. 
In Fig.~\ref{fig:LFVNHIH} we display the dependence of the three observables $\mathcal{B}(\mu\rightarrow e \gamma)$ (green points), $\mathcal{R}_{\mu e}$ (cyan cross) and $\mathcal{B}(\mu\rightarrow 3e)$ (red dots) with $\sqrt{|\rho_1||\rho_2|}$ -the Yukawa couplings that are unconstrained by neutrino physics- (left panels) and with $\eta_1$ -the only free Yukawa coupling that enter in the neutrino mass generation- (right panels). 
From the results for the NH, the upper bounds $\sqrt{|\rho_1||\rho_2|} \lesssim 10^{-1}$ and $|\eta_1|\lesssim 10^{-2}$ are obtained.  
Note that the most constraining limits are those from $\mu-e$ conversion experiments, which will test Yukawa couplings down to $5\times10^{-3}$.  
On the other hand, the results for inverted hierarchy give the upper bounds $|\eta_1|\lesssim 10^{-1}$ and $\sqrt{|\rho_1||\rho_2|} \lesssim 10^{-1}$, which may improve in around one order of magnitude in the future.
The lower bound on $|\eta_1|$ that appears in the bottom right panel of  Fig.~\ref{fig:LFVNHIH} is due to the fact that for the IH the magnitude of the $\eta_{2,3}$ is less that $|\eta_1|$, which implies that for $|\eta_1|\approx10^{-5}$ that value is excluded due to the lower limit imposed over all the Yukawa couplings. 
It is worth mentioning that the results for the three Yukawa  couplings $\rho_i$ are similar for both neutrino mass spectrum since these couplings do not participate in the neutrino mass generation.

\begin{figure}[t!]
  \begin{center}
    \includegraphics[scale=0.45]{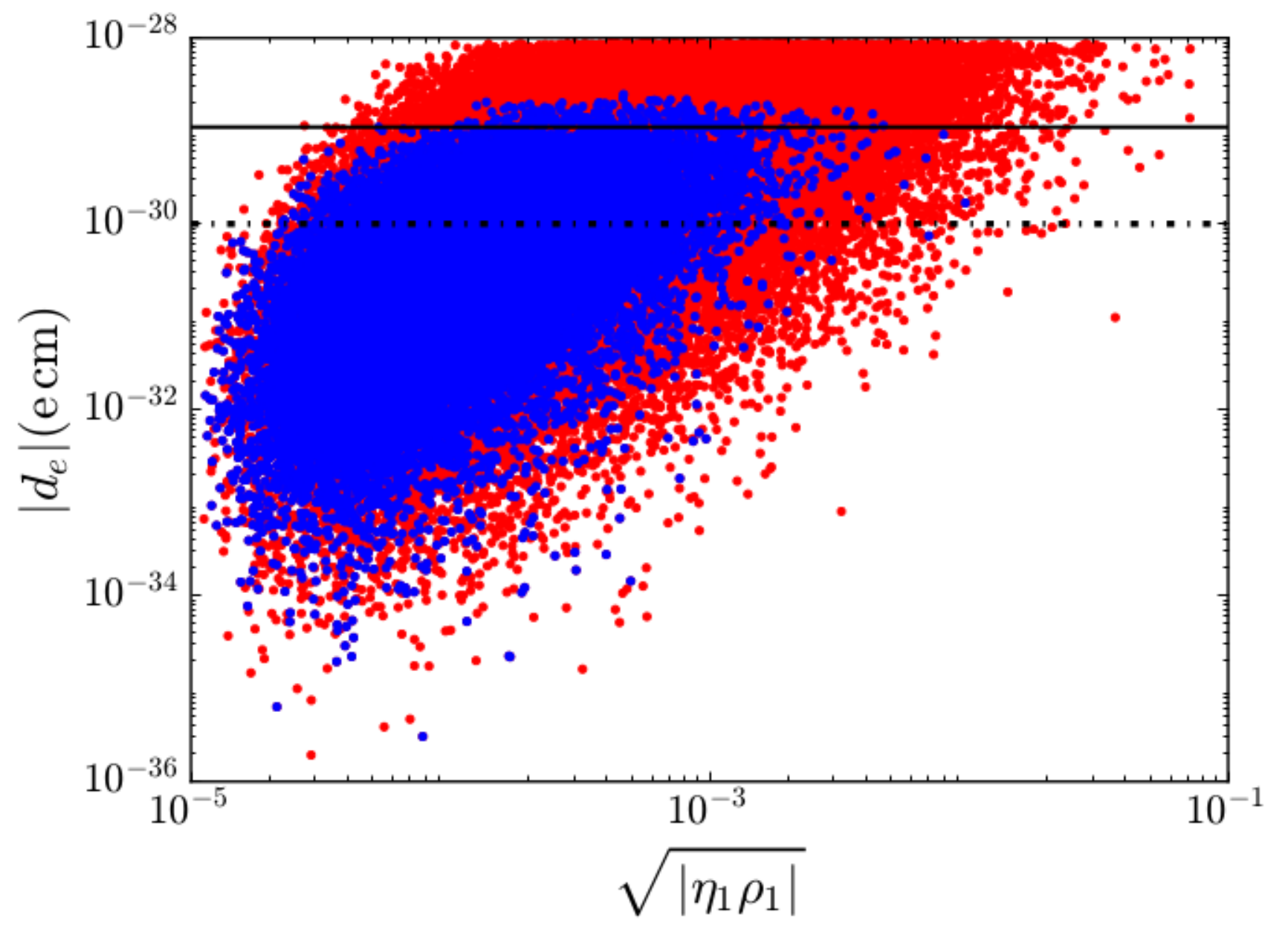}
    \includegraphics[scale=0.45]{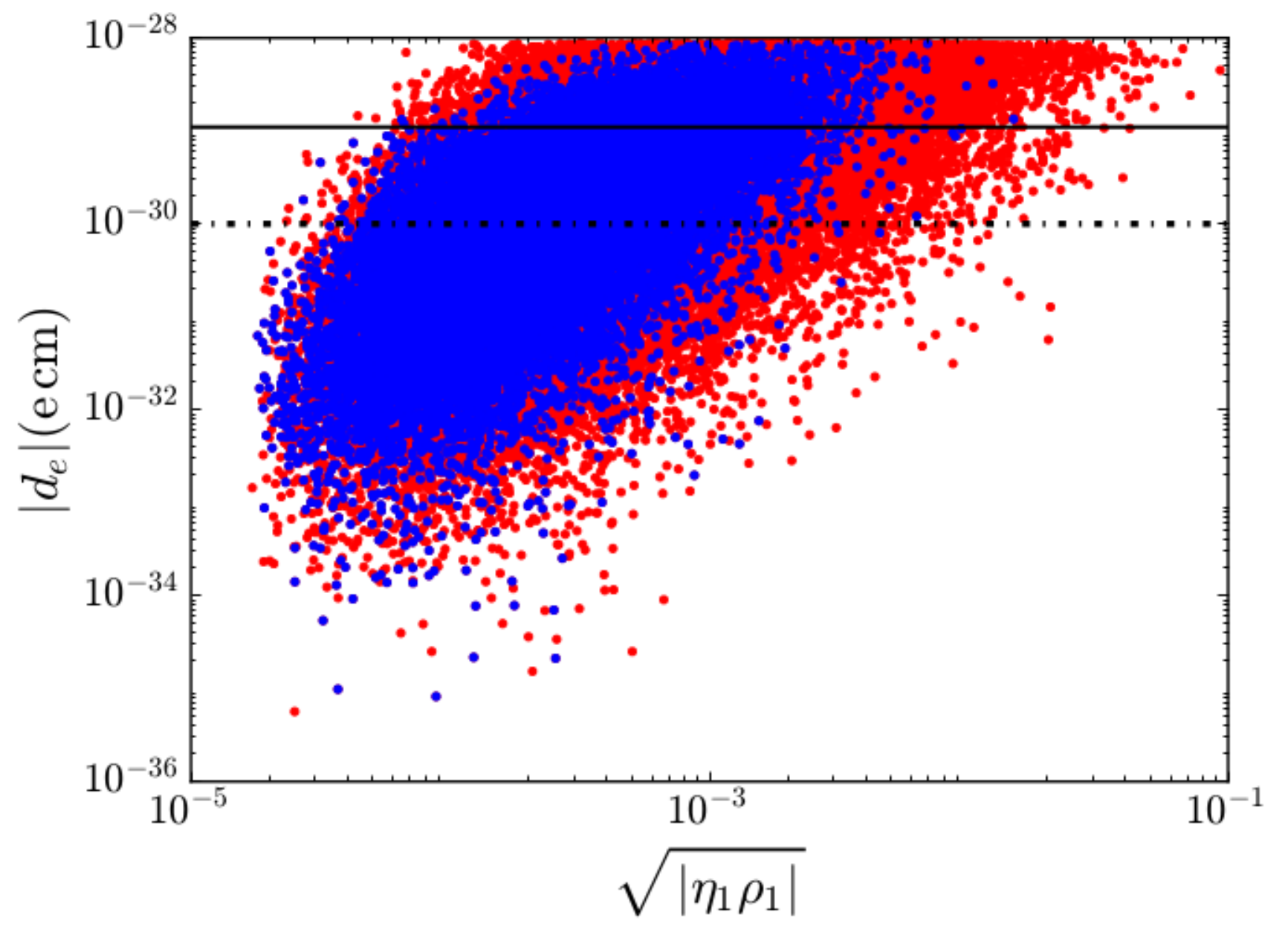}
    \caption{Expected values for the eEDM as a function of $\sqrt{|\eta_1\rho_1|}$ for NH (left panel) and IH (right panel).
      The red points constitute the current viable parameter space while the blue points are beyond the reach of future CLFV searches.
    }
    \label{fig:EDM1}
  \end{center}
\end{figure}

In Fig.~\ref{fig:LFV1} we show the correlation between the $\mathcal{B}(\mu\rightarrow e \gamma)$, $\mathcal{B}(\mu\rightarrow 3e)$ and $\mathcal{R}_{\mu e}$ observables, and the impact of the future searches associated to these observables over the parameter space considered. The correlation between the three observables is remarkable in a large portion of the parameter space. 
It follows that a large fraction of the current viable parameter space will be tested in the future experiments, with all the electron-muon observables  being within the reach of the MEG and Mu3e experiments, and with the $\mathcal{R}_{\mu e}$ observable being the most promising. 

The results for the eEDM as a function of the Yukawa coupling product $\sqrt{|\rho_1\eta_1|}$ are displayed in Fig.~\ref{fig:EDM1}.
Our results show that eEDM future searches \cite{Baron:2013eja} (dashed lines) may test regions beyond the reach of experimental sensitivity of CLFV searches. 
Indeed, the most recent result on the eEDM from ACME collaboration \cite{Andreev:2018ayy} has begun to test those regions (solid lines in Fig.~\ref{fig:EDM1}).
It turns out that these regions are precisely formed by benchmark points (for both neutrino mass spectrum) with $\mathcal{B}(\mu\to 3e)/\mathcal{B}(\mu\to e\gamma)>1$ and $\mathcal{R}_{\mu e}(\text{Ti})/\mathcal{B}(\mu\to e\gamma)\gtrsim10$, that is, where the dipole contribution is not the dominant one.
This means that, if it is experimentally observed that the above ratios are below the mentioned bounds, the parameter space we are considering would be disfavoured. 

\begin{figure}[t!] 
\includegraphics[scale=0.42]{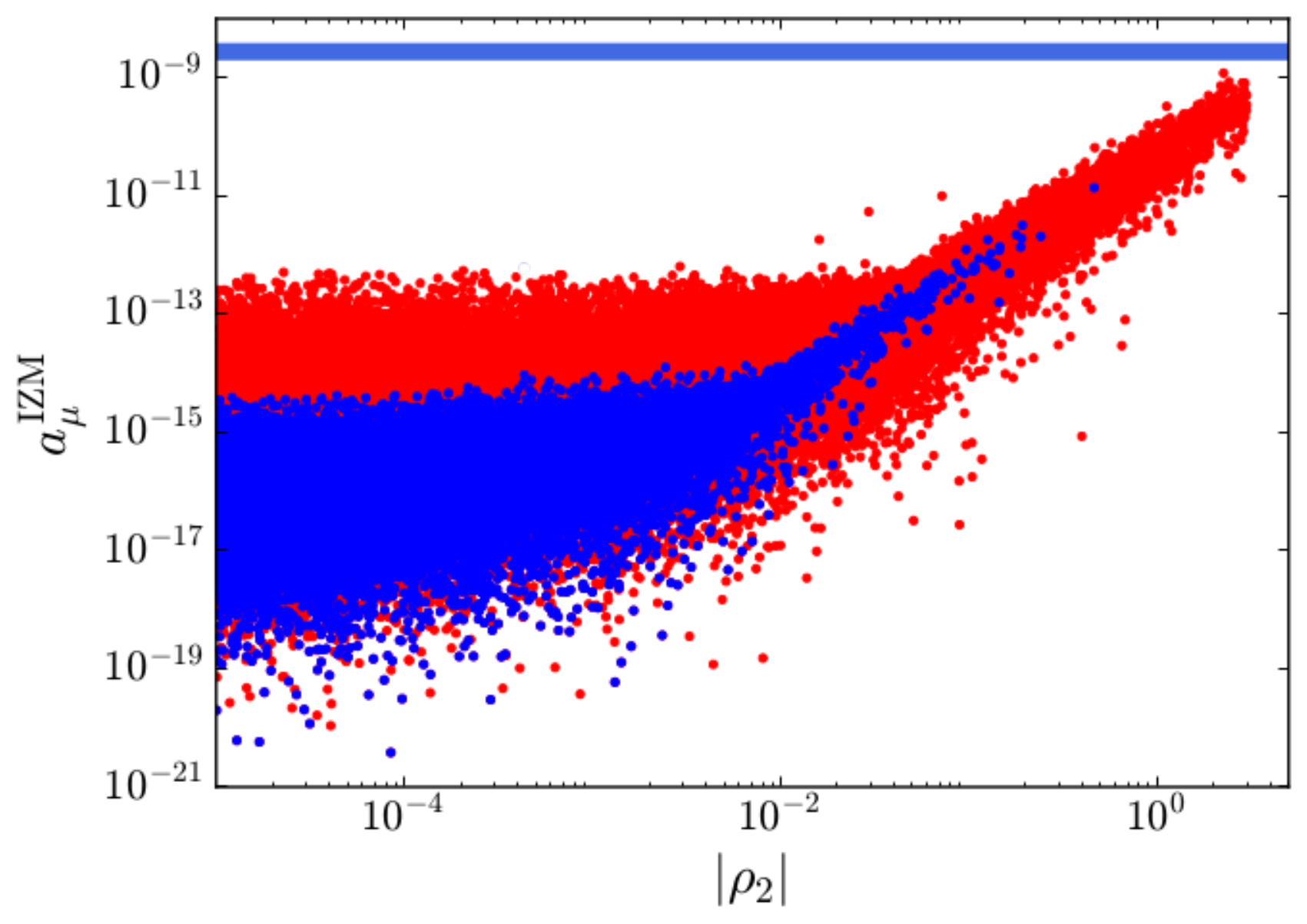}
\includegraphics[scale=0.42]{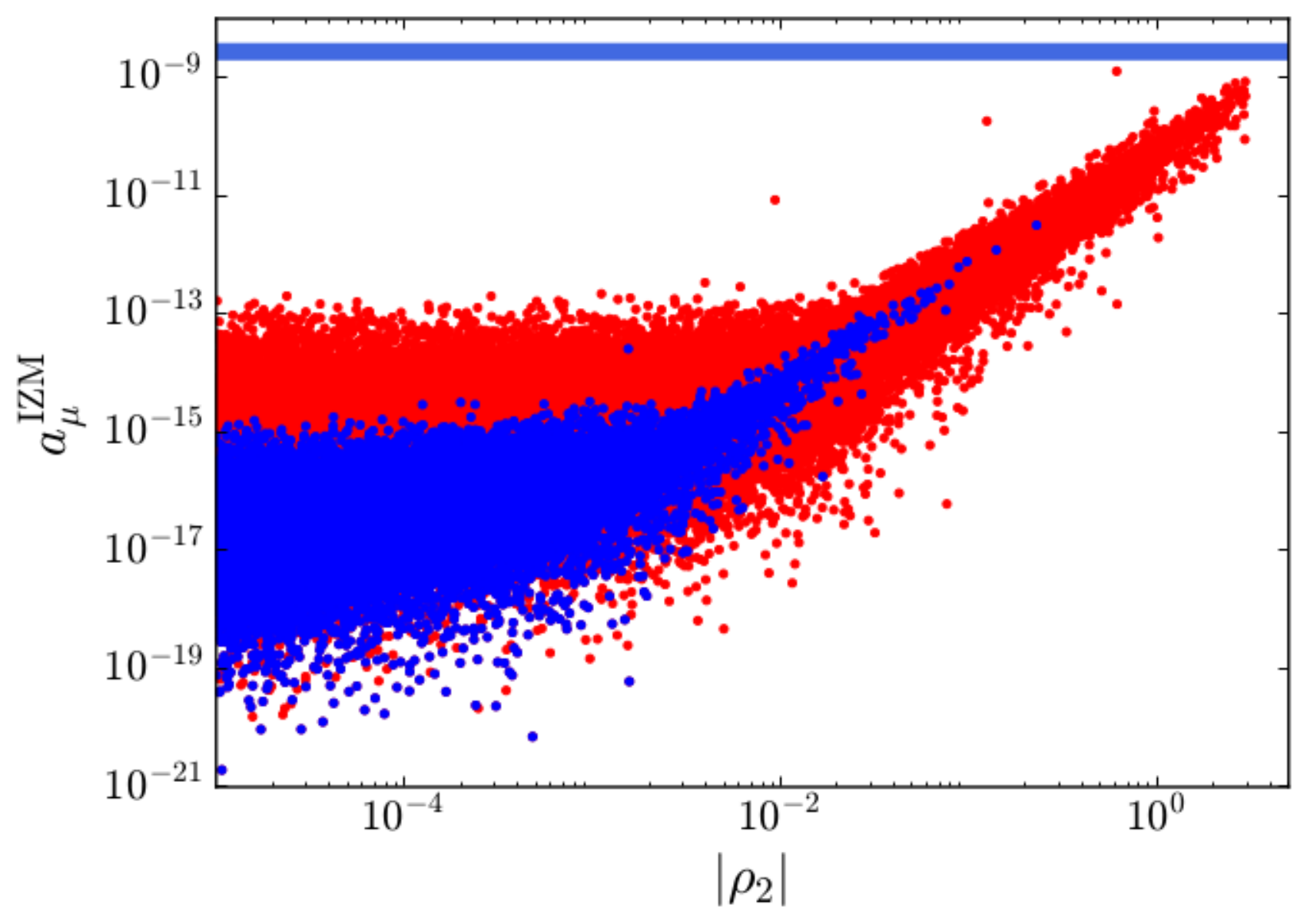}\\
\caption{The new contribution $a_\mu^{\text{IZM}}$ to the anomalous magnetic dipole moment of the muon  as a function of $|\rho_2|$ for NH (left panel) and IH (right panel). The horizontal blue band represents the discrepancy between the SM prediction and the experimental value \cite{Lindner:2016bgg}. 
}
\label{fig:amu}
\end{figure}

Finally, the contribution to the anomalous magnetic moment of the charged leptons are not so relevant since they are below the current bounds \cite{Patrignani:2016xqp}. 
For completeness purposes we display in Fig.~\ref{fig:amu} the results for the anomalous magnetic moment of the muon.
In particular, the new contribution is not enough to explain the discrepancy between the SM prediction and the experimental value $\Delta a_\mu=a_\mu^{\text{exp}}-a_\mu^{\text{SM}}=(288\pm80)\times10^{-11}$ \cite{Bennett:2006fi,Patrignani:2016xqp}. 

\section{Conclusions}
\label{sec:conclusions}
We have explored the inert Zee model in  light of the ambitious experimental program designed to probe, via charged LFV processes and EDM signals, beyond the Standard Model scenarios. 
We determined the viable parameter space consistent with the current constraints coming from a diversity of directions: dark matter, neutrino oscillations, lepton flavor violating processes, electric dipole moments, electroweak precision tests and collider physics.
We have also established the most relevant experimental perspectives regarding LFV searches, where we have found that $\mu-e$ conversion in muonic experiments constitutes the most promising way in this line of research.  
Furthermore, since the Yukawa couplings that reproduce the neutrino oscillation observables are complex, which in turn provide new sources of CP violation, we have shown the regions in the parameter space where the prospects for the eEDM are within the future experimental sensitivity.
It is remarkable the impact that may have eEDM future searches since they may probe the model in regions out the reach of all the future CLFV projects.  

\section*{Acknowledgments}
We are grateful to Federico von der Pahlen for comments about the manuscript. 
This work has been partially supported by the Sostenibilidad program of Universidad de Antioquia UdeA, CODI-E084160104 grant and by COLCIENCIAS through the grants 111565842691 and 111577657253. O.Z. acknowledges the kind hospitality of the Abdus Salam ICTP (through the Simons Associateships) where the final part of this work was carried out.

\appendix{}

\section{Loop functions}
\label{sec:Loopmuegama}
The analytical expressions for the loop functions involved in the $d_{\ell_i}$, $a_{\ell_i}$ and $\ell_i\to \ell_j\gamma$ observables are:
\begin{align}
\mathcal{G}_1(m_a^2,m_b^2,m_c^2)&=\frac{1}{m_b^2} G\left(\frac{m_a^2}{m_b^2}\right)-\frac{1}{m_c^2} G\left(\frac{m_a^2}{m_c^2}\right),\\
\mathcal{F}_1(m^2_{a},m_{b}^2,m_{c}^2)&=\frac{1}{2 m_{a}^2} \left[ F\left(\frac{m_{b}^2}{m_{a}^2}\right)+ F\left(\frac{m_{c}^2}{m_{a}^2}\right) \right],\\
\mathcal{F}_2(m^2_{a},m_{b}^2)&=\mathcal{F}_1(m_{a}^2,m_{b}^2,m_{b}^2), \\
\mathcal{I}_1 (m_{a}^2, m_{b}^2,m_{c}^2 ) &= \frac{1}{m_a^2} \left[ G_2\left(\frac{m^2_{b}}{m^2_{a}}\right) -  G_2\left(\frac{m^2_{c}}{m^2_{a}}\right) \right],\\
\mathcal{H}_1 (m_{a}^2, m_{b}^2,m_{c}^2 ) &=  \frac{1}{m_a^2} \left[ F\left(\frac{m^2_{b}}{m^2_{a}}\right)-F\left(\frac{m^2_{c}}{m^2_{a}}\right) \right],\\
\mathcal{H}_2 (m_{a}^2, m_{b}^2) &= \mathcal{H}_1 (m_{a}^2, m_{b}^2,0 ) =  \frac{1}{m_a^2}  F\left(\frac{m^2_{b}}{m^2_{a}}\right),
\end{align}
where
\begin{align}
 F\left(x\right) &= \frac{2x^3 + 3x^2 -6x +1 -6x^2\log\left(x\right)}{6\left(x-1\right)^4},\\ 
 G\left(x\right) &= \frac{x^2 -4x + 3 + 2\log\left(x\right)}{2\left(x-1\right)^3},\\
  G_2(x) &= \frac{3x^2-4x +1 -2 x^2 \mbox{log}(x)}{2(1-x)^3}.
\end{align}

\bibliographystyle{h-physrev4}
\bibliography{darkmatter}
\end{document}